\documentclass[aps,pra,twocolumn]{revtex4}

\usepackage[dvips]{graphicx}
\usepackage{amsmath}

\begin{document}

\title{Two-dimensional transport and transfer of a single  atomic qubit in optical tweezers}

\author{J. Beugnon, C. Tuchendler, H. Marion, A. Ga\"{e}tan, Y. Miroshnychenko, \\
Y.R.P. Sortais, 
A.M. Lance M.P.A. Jones, G. Messin, A. Browaeys, P. Grangier}
\affiliation{Laboratoire Charles Fabry de l'Institut d'Optique, CNRS, Univ. Paris-sud,
Campus Polytechnique,
RD 128, 91127 Palaiseau cedex, France}
\date{\today}


\maketitle

{\bf Quantum computers have the capability of out-performing their classical 
counterparts for certain computational problems~\cite{Nielsen}. 
Several scalable quantum computing architectures 
have been  proposed. An attractive architecture is a large set of physically 
independant qubits, arranged in three spatial regions where 
(i) the  initialized qubits are stored in a register, 
(ii) two qubits are brought together to realize a gate, and 
(iii) the readout of the qubits is performed~\cite{kielpinski02,Calarco04}.  
For a neutral atom-based architecture, a natural way to connect 
these regions is to use optical tweezers to move qubits within the system.
In this letter we demonstrate the coherent transport of a qubit, 
encoded on an atom trapped in a sub-micron tweezer, over a distance typical of the separation 
between atoms in an array of optical traps~\cite{Dumke02, Bergamini04, Miroschnychenko06}. 
Furthermore, we transfer a qubit between two tweezers, and 
show that this manipulation also preserves the  coherence of the qubit.}

In the quest for an implementation of a quantum computer, scalability is a major concern.   
In the trapped ion approach (see e.g.~\cite{QFFT}), a lot of effort is being devoted 
to building arrays of small ion traps~\cite{Seidelin06}, and to moving  
ion qubits whilst avoiding  heating and decoherence~\cite{Rowe02}. 
Neutral atoms also offer promising properties for the realization of large quantum registers.  
For example, one- or two-dimensional adressable arrays of dipole traps have 
been demonstrated using holographic techniques~\cite{Bergamini04}, 
micro-fabricated elements~\cite{Dumke02}, or active rearrangement 
of single atoms~\cite{Schrader04, Miroschnychenko06}. An
 alternative approach is to use the Mott insulator transition to 
 initialize a three-dimensional register by loading a Bose-Einstein 
 condensate into an optical lattice~\cite{Greiner01}. 
 Recent progress has shown subwavelength addressability in such a system~\cite{Lee07}.  
 To perform quantum computations, however, an additional key feature is the ability to 
 perform the gate between two arbitrary qubits of the register.

Here we demonstrate a scheme where a neutral atom qubit is 
transfered between two moving tweezers (``register"  to ``moving head"), 
and then transported towards an interaction zone where the two-qubit 
gate should be implemented~\cite{jaksch99,Brennen99,jaksch00,Dorner05}.  
We show that these manipulations of the external degrees of 
freedom preserve the coherence of the qubit, and do not induce any heating. 
This transport in a moving tweezer is a promising alternative to the recently 
demonstrated transport of qubits in ``optical conveyor belts"~\cite{Kuhr03,Miroschnychenko06}, 
or in state-dependent moving optical lattices~\cite{Mandel03a}. 
Altogether, these results pave the way towards a scalable neutral atom quantum computing architecture.

In our experiment, 
we trap a single rubidium 87 atom in an optical dipole trap created by a tightly focused laser beam~\cite{Schlosser02,Sortais07}. 
As described in detail in~\cite{Jones07}, the qubit is encoded onto the $|0\rangle=|F=1,M=0\rangle$ and $|1\rangle=|F=2,M=0\rangle$ 
hyperfine ground states separated by $\omega_{\rm{hf}}\approx 6.8$ GHz.
We initialize the qubit in state $|0\rangle$ by optically pumping the atom.
We drive single-qubit operations by a Raman transition using two phased-locked laser beams, one 
of which being the dipole trap. 
The internal dephasing time of the qubit, measured  by Ramsey interferometry, is  $\approx 630\ \mu$sec. 
This time is mostly limited by the residual 
motion of the atom in the trap that leads to a 
fluctuation of the frequency of the qubit transition.
This dephasing can be reversed by applying a spin echo technique where a 
$\pi$ pulse is inserted between the
two $\pi/2$ pulses of the Ramsey sequence. 
Using this technique we measure an irreversible dephasing time of 34 ms.

\begin{figure}
\includegraphics[width=7 cm]{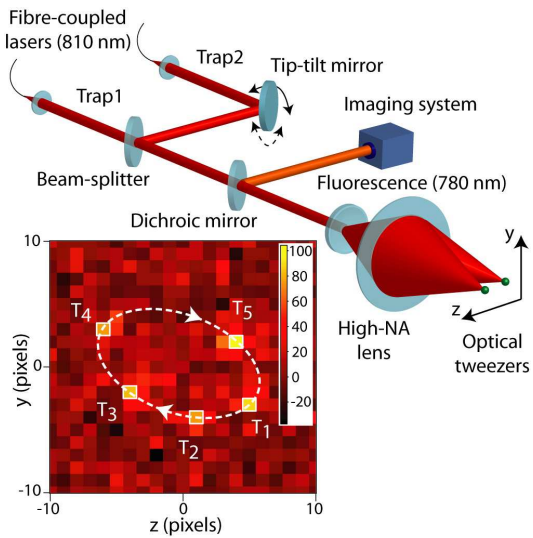}
\caption{Experimental setup. A large numerical aperture lens focuses two independent 
dipole trap beams at 810 nm each to a size of 0.9 micrometers.
An optical power of 400 $\mu$W results into a trap depth of 500 $\mu$K and oscillation frequencies of 81 kHz and 15 kHz, in the 
radial and axial directions respectively.
The two trapping lasers have the same linear polarization and their frequencies are separated by 10 MHz to avoid interferences. 
The moving tweezer is displaced by rotating a tip-tilt platform. 
The same large numerical aperture lens is used to collect the fluorescence light 
at 780 nm from the atom. This fluorescence light is separated from the trapping light by the dichroic mirror and sent to a 
single-photon counter module and a CCD camera. The insert shows a fluorescence picture of an atom moved along an elliptical 
trajectory
in the y-z plane. The picture is a summation of 5 images taken at different times during the motion. }
\label{experimentalsetup}
\end{figure}

The experimental setup for the moving tweezer is represented in figure~\ref{experimentalsetup}.
We reflect the dipole trap beam off a mirror affixed to a tip-tilt platform 
prior to the large numerical aperture lens. The platform is actuated by piezo-electrical transducers and can 
rotate with a maximal angle of 2.5 mrad in both the horizontal and vertical directions.
We have measured the position of the dipole trap for different 
angles of the platform by observing the position of the atom on the CCD camera. 
The maximal angle corresponds to a total displacement of the tweezer of $18\pm 1$ $\mu$m. This motion 
is two-dimensional, as demonstrated in figure~\ref{experimentalsetup}.

We first analyze the influence of a displacement of the tweezer 
on the external degrees of freedom of the qubit. For this purpose, we measure the temperature of the 
single atom in the tweezer using a release and 
recapture technique~\cite{Lett88} (see Methods). In the absence of motion, 
the temperature of the atom is $56.0\pm 1.4$ $\mu$K. We repeat this measurement 
after moving the tweezer by a total distance of 360 $\mu$m, 
consisting of 20 round trips of 18~$\mu$m, along the $z$-axis.
Each round trip lasts a time of 6 ms. We do not find any measurable loss due to this transport. 
After the motion, we measure a temperature of 
 $54.8\pm 1.6$ $\mu$K. As the energy difference between two vibrational quanta in the radial 
 direction
 is 4 $\mu$K, this temperature is compatible with no change in the radial vibrational state. 
This absence of motional heating  is a crucial feature for entanglement schemes based on controlled 
collisions~\cite{Dorner05,Mandel03b} and 
results from the adiabaticity of the displacement.
A motion is adiabatic if the acceleration $a$ fulfills $m a \sigma \ll \hbar \Omega$ ($\Omega$ is 
the oscillation frequency of the atom, $m$ its mass and $\sigma$ the extension 
of the ground state wave function~\cite{browaeys06}). This gives a maximum acceleration 
of $\approx 10^4$  m/$\rm{sec}^2$, much larger than the experimentally measured $\approx 15$ m/$\rm{sec}^2$.

\begin{figure}
\includegraphics[width=6 cm]{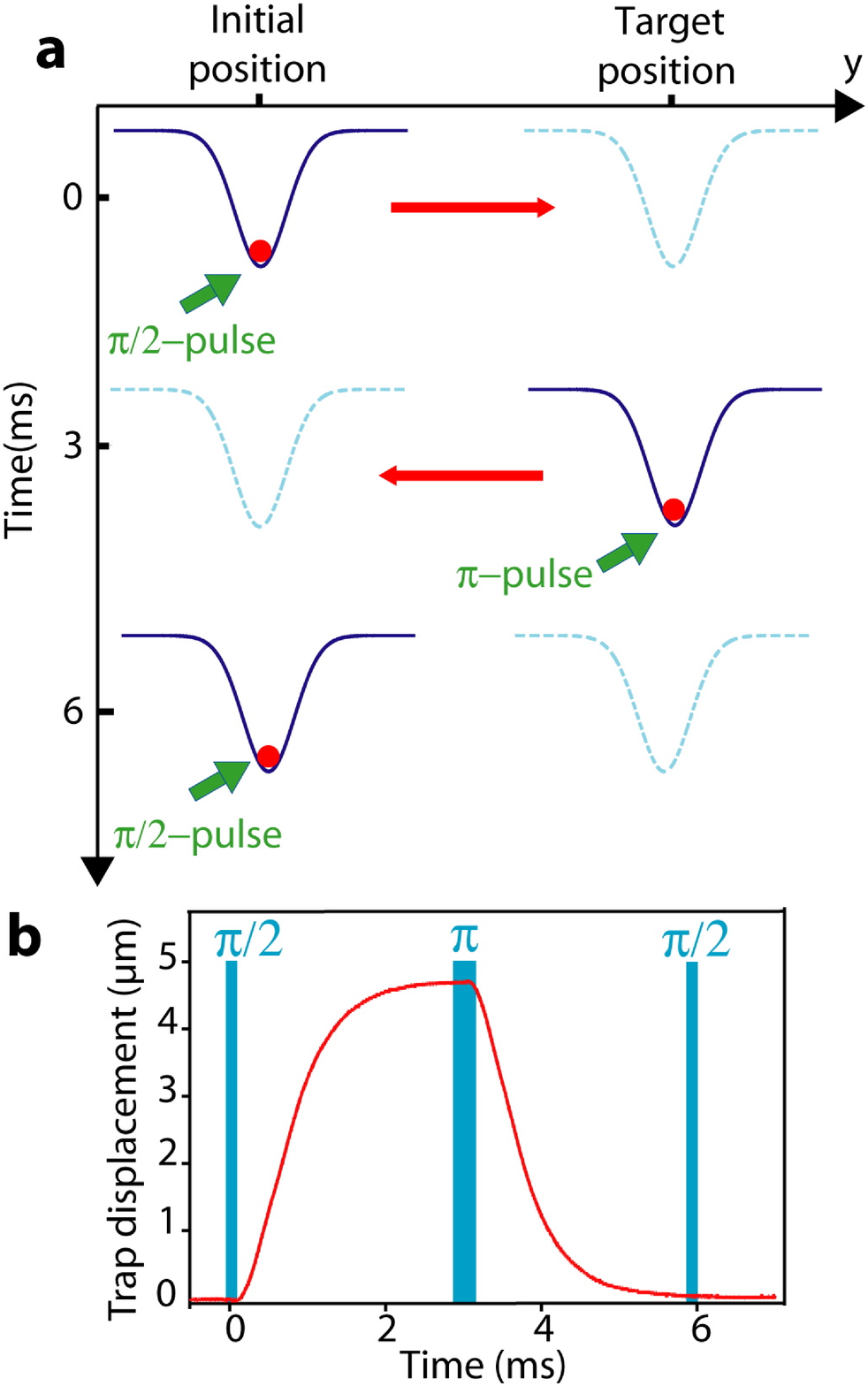}
\caption{Principle of the moving qubit experiment. (a)
Position of the atom when the pulses of the 
spin echo sequence are applied. Starting at the initial position, the first pulse prepares the atom in a 
superposition $(|0\rangle + |1\rangle)/\sqrt{2}$. 
The tweezer is then moved  along the $y$-axis to the target position and
the $\pi$ pulse is applied. Finally, the tweezer is brought back to its initial 
position and the coherence is checked by applying a second $\pi/2$ pulse and measuring
the state of the qubit. (b) Example of displacement of the tweezer versus time. 
The signal is obtained from the sensor attached to the tip-tilt
platform, which is converted into a distance travelled by the tweezer. The $\pi/2$ and $\pi$  
pulses, being 2 and 4 $\mu$s long respectively, are not to scale.}
\label{principemove}
\end{figure}

\begin{figure}
\includegraphics[width=7 cm]{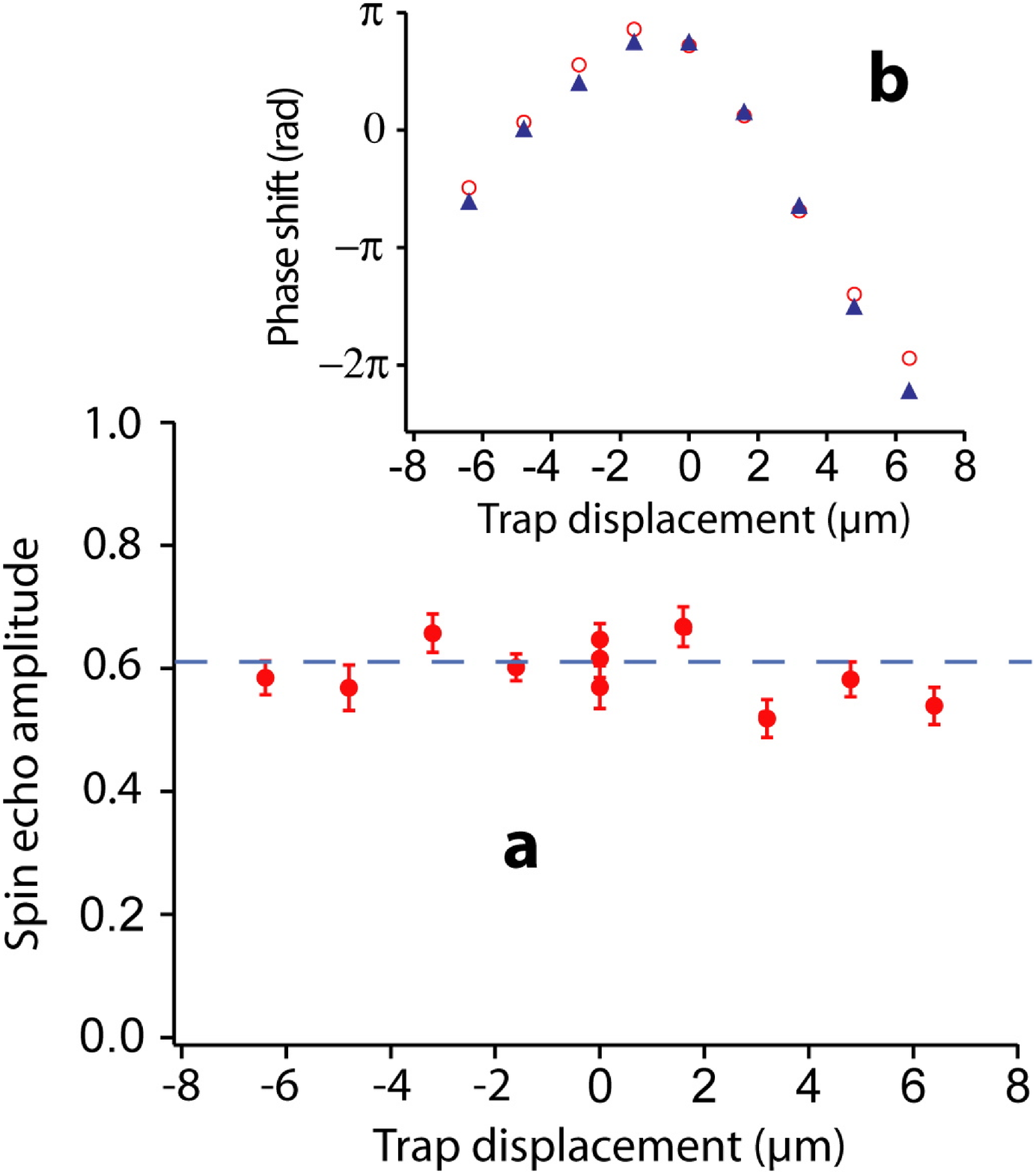}
\caption{Results of the moving qubit experiment.  Figure (a) shows the amplitude
of the spin echo signal versus the amplitude of the displacement. The error bars are the root mean square (RMS) uncertainty obtained
from the fit of the fringes. The 60\% contrast is a result of the damping of the fringes after 6 ms~\cite{Jones07}. The dashed line 
represents the average of the data for no trap displacement.
Figure (b) shows the phase shift versus the amplitude of the 
displacement. The triangles are the data and the circles are the calculated value of the dephasing based on the model 
described in the Methods section.}
\label{movingqubit}
\end{figure}

We secondly study the influence of the motion on the coherence of the qubit. 
As the duration of the transport is larger than the dephasing time of the qubit (630 $\mu$s), 
we  apply the spin echo sequence to rephase the qubit~\cite{Jones07}. The time sequence of the experiment 
is shown in figure~\ref{principemove}. 
Figure~\ref{movingqubit}(a) shows the amplitude of the spin-echo fringes for various 
trap displacements along the $y$-axis. This amplitude is  constant when we scan the tweezer over all the transverse field 
of the objective. This demonstrates that the motion does not affect the internal
coherence of the qubit. We observe the same behavior when we move the tweezer  along the $z$-axis.

We also observe a phase shift of the spin-echo fringes, as shown in figure~\ref{movingqubit}(b). This 
is a signature that the two states of the qubit dephase with respect to each other during the motion 
(in a reproducible way), despite the 
presence of the rephasing $\pi$ pulse. We attribute this phase shift to the  asymmetry 
of the trajectory during the first and second part of the 
round trip displacement. We have modelled this effect  and found a  good
agreement with the data (see Methods section).  This understanding of the phase evolution of the qubit during the motion
is crucial for a possible implementation in a quantum computer where qubit phases need to be controlled.

\begin{figure}[h]
\includegraphics[width=6 cm]{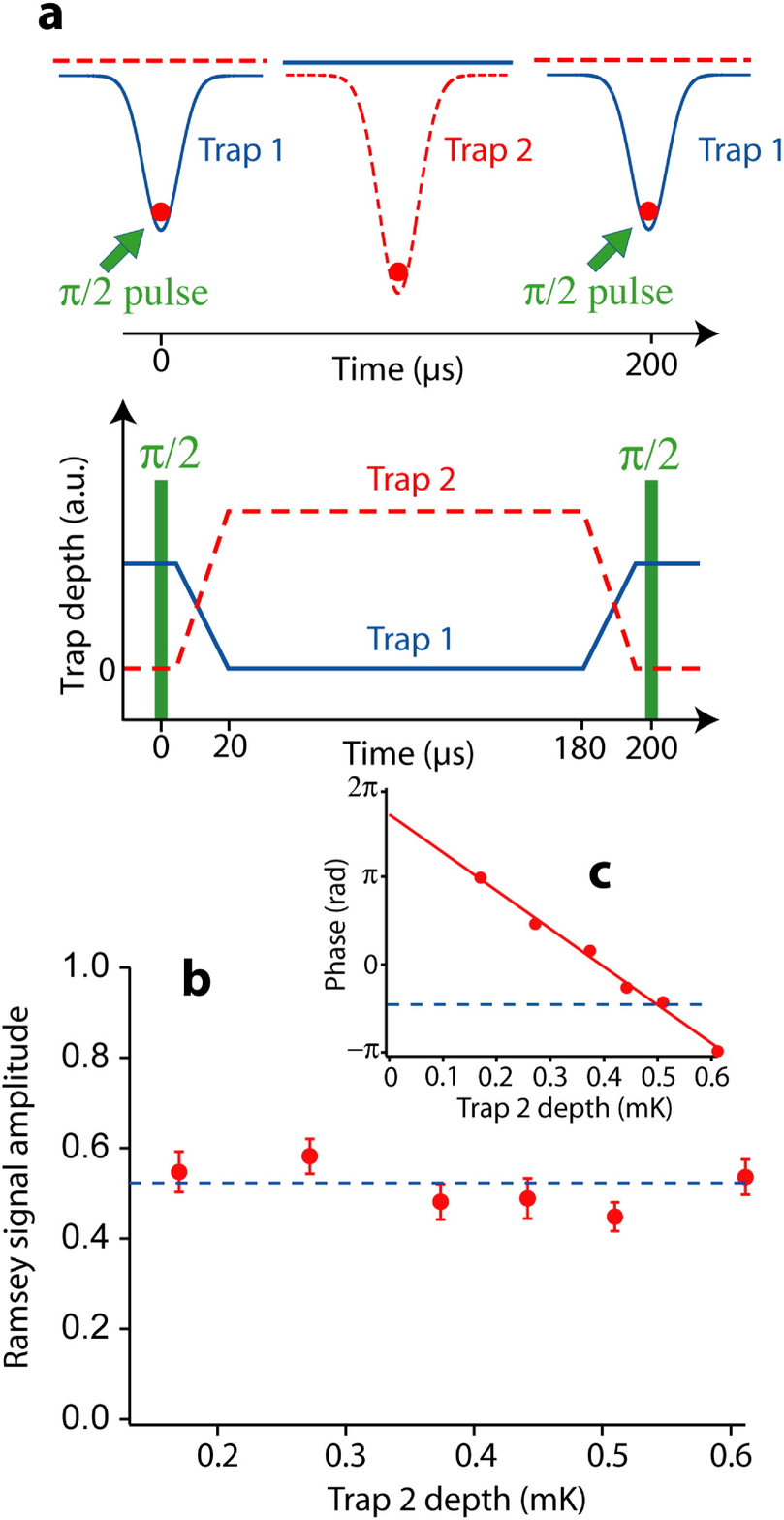}
\caption{Experiment on the transfer of the qubit between two tweezers.
Figure (a) details the time sequence. A $\pi/2$ pulse is applied when the atom is initially in trap 1, which has a depth of 
500~$\mu$K. The atom is transferred 
to trap 2 in a time of 20 $\mu$s. After a time of 160~$\mu$s, the atom is transferred back to trap 1 and 
a second $\pi/2$ pulse is applied.
Figure (b) shows the amplitude of the spin echo signal as a function of the depth of the trap the atom is transferred into. 
The error bars are the RMS uncertainty obtained
from the fit of the fringes.
Figure (c) displays the phase of the Ramsey oscillations for different depths of the second trap. The  solid line is a linear fit to the data.
The dashed lines in (b) and (c) correspond to  no transfer. }
\label{transferqubit}
\end{figure}

With the idea of transfering an atom from the ``register'' to the ``moving head'', 
we have investigated the transfer of a qubit from one tweezer to a second one. For this experiment,
the two traps are superimposed and the positions of both tweezers are fixed.
The experimental sequence is shown in figure~\ref{transferqubit}(a). We load an atom in the first 
tweezer, transfer it to a second tweezer, and transfer it back to the first tweezer, with no measurable loss. 
When the two traps have the same depth, we measure a temperature of the atom  after the double transfer of 
$56.3\pm 1.8$ $\mu$K, while the  temperature with no transfer is $53.4\pm1.4$ $\mu$K.
Therefore, the transfer does not induce any significant motional heating.

We analyze the influence of the transfer on the coherence of the qubit 
by inserting the double transfer between the two $\pi/2$ pulses
of a Ramsey sequence, as shown in figure~\ref{transferqubit}(a).
Figure~\ref{transferqubit} presents the amplitude  and the phase of the Ramsey fringes after this sequence 
for depths of the second tweezer ranging from 0.2 mK to 0.6 mK. 
This transfer  does not affect the amplitude of the Ramsey signal when the depth of the second trap is varied, thus 
showing that the coherence is robust against the transfer between the two traps. 

Figure~\ref{transferqubit}(c) shows that the phase of the Ramsey fringes,
varies linearly with respect to the  depth of the second trap.
This is explained by the differential potential experienced by the two states, which is proportional to the depth of the trap $U$.
If the depths of the two traps differ
by $\Delta U$, the Ramsey fringes are shifted after a holding time $T$ 
by a phase proportional to $\Delta U\,  T/ \hbar$, with respect to the situation where no transfer is applied.
This phase is thus a useful tool to make sure that the two traps are identical.  

As a conclusion we have shown that we can move and transfer a single qubit between two tweezers
with no measurable motional heating. 
We have also shown no loss of coherence of the atomic qubit under transfer and displacement. 
In combination with our holographic array of dipole traps~\cite{Bergamini04},  and  efficient  
single-qubit operation and readout~\cite{Jones07}, we have made a first step towards 
designing a scalable architecture of a quantum computer based on neutral atoms.

\vspace{5mm}

{\bf Methods:}

\vspace{2mm}

Temperature measurement:

\vspace{2mm}
After trapping a single atom, we switch off the dipole trap for a time adjustable between 1 and $\sim 30$ $\mu$sec. 
We then turn the 
trap back  on  and check for the presence of the atom. 
We repeat this sequence 100 times for each release time and 
calculate the probability to recapture the atom after the corresponding time of flight. 
We compare our data with a 3-dimensional Monte Carlo 
simulation, taking into account the potential produced by the Gaussian trapping beam,
and assuming a thermal distribution of the position and the velocity of the atom at the beginning of the time of flight.
The error bar of this fitted temperature corresponds to one standard deviation in the least square-based fit.

\vspace{2mm}
Phase shift during the motion:
\vspace{2mm}

The hyperfine splitting of 6.8 GHz means that the dipole trap detuning 
is slightly larger for $|0\rangle$ than for $|1\rangle$, giving rise to a small differential lightshift.
Therefore the qubit transition frequency is $\omega_{\rm{hf}} + \eta U/\hbar$,
with $U$ the depth of the dipole trap and $\eta=7 \times 10^{-4}$  for our trap. If the tweezer is not moved,
the $\pi$-pulse compensates for the phase accumulated during the two parts of the motion.  
When the tweezer is moved off axis, the waist of the 
beam increases sligthly, resulting in a shallower trap. 
Figure~\ref{principemove}(a) shows that with the tweezer starting on axis, 
the atom spends more time far from the axis where
the dipole trap is shallower, whereas on the way back it spends more time around the axis 
where the dipole trap is stronger. The average depth is then different for the two parts of the motion, and
so are the phases.  As the  phase of the spin-echo signal is the 
difference of the phases accumulated during the two periods of the motion, 
it is expected to vary as we move the tweezer further away
off axis. We have modelled this effect by calculating the dephasing accumulated 
during the transport, taking into account 
the actual displacement of the tweezer from the sensor curve and the measured Rabi frequencies  for 
different positions of the tweezer off axis. The result of this model is shown as circles in figure~\ref{movingqubit}(b) 
and is consistent with the data.
  
{\bf Acknowledgments:}
We would like to thank W.D. Phillips, T. Porto, I. Deutsch, P. Jessen for stimulating discussions.
We acknowledge financial support from 
IFRAF, ARDA/DTO and the European Integrated project SCALA. LCFIO is  CNRS UMR8501. 
M.P.A. Jones and A.M. Lance are supported by Marie Curie Fellowships. A. Ga\"{e}tan is supported by a DGA Fellowship.

\end{document}